\documentclass[aps,preprintnumbers, superscriptaddress, graphicx, epsfig, amsmath, showpacs]{revtex4}
\usepackage{graphicx}
\usepackage{psfrag}

\usepackage{graphicx}
\usepackage{dcolumn}
\usepackage{bm}
\usepackage{epstopdf}

\begin{document}


\title{New look at the QCD ground state in a magnetic field}

\author{Efrain J. Ferrer, Vivian de la Incera, Israel Portillo  and Matthew Quiroz\\\textit{Department of Physics, University of Texas at El Paso, El Paso, Texas
79968, USA}}

\date{\today}

\begin{abstract}
We explore chiral symmetry breaking in a magnetic field within a Nambu-Jona-Lasinio model of interacting massless quarks including tensor channels. We show that the new interaction channels open up via Fierz identities due to the explicit breaking of the rotational symmetry by the magnetic field. We demonstrate that the magnetic catalysis of chiral symmetry breaking leads to the generation of two independent condensates, the conventional chiral condensate and a spin-one condensate. While the chiral condensate generates a dynamical fermion mass, the new condensate gives rise to a dynamical anomalous magnetic moment for the fermions. As a consequence, the spectrum of the excitations in all Landau levels, except the lowest one, exhibits Zeeman splitting. Since the pair, formed by a quark and an antiquark with opposite spins, possesses a resultant magnetic moment, an external magnetic field can align it giving rise to a net magnetic moment for the ground state. This is the physical interpretation of the spin-one condensate.  Our results show that the magnetically catalyzed ground state in QCD is actually richer than previously thought. The two condensates contribute to the effective mass of the LLL quasiparticles in such a way that the critical temperature for chiral symmetry restoration becomes enhanced.
\end{abstract}



\maketitle

\section{Introduction}
Understanding the phases of matter under strong magnetic fields constitutes an active topic of interest and debate in light of contradictory theoretical results about the influence of a magnetic field in the chiral and deconfinement transitions of quantum chromodynamics (QCD) \cite{CSB}-\cite{QCD-latticeB}; as well as due to the existence of large magnetic fields in compact stars and their production in heavy-ion collisions.

Extremely high magnetic fields $eB\approx 2m_\pi^2$ ($\sim 10^{18}$ G) \cite{Heavy-Ion-1}-\cite{Heavy-Ion-2} can be generated in noncentral Au-Au collisions for top collision energies $\sqrt{S_{NN}}=200$ GeV at the Relativistic Heavy Ion Collider (RHIC) at Brookhaven National Lab (BNL). Even though these magnetic fields decay quickly, they only decay to a tenth of the original value for a time scale of order of the inverse of the saturation scale at RHIC \cite{Tuchin}-\cite{Fukushima-Pawlowski}, hence they may influence the properties of the particles generated during the collision. Even larger fields, of order $eB\approx 15m_\pi^2$ ($\sim 10^{19}$ G), can be generated for the energies reachable at the Large Hadron Collider (LHC) at CERN, $\sqrt{S_{NN}}=4.5$ TeV, for the Pb-Pb collisions \cite{Heavy-Ion-2}. Later in this decade, the Facility for Antiproton and Ion Research (FAIR) at GSI will open the possibility to explore the intermediate region of temperatures and densities, thereby expanding our understanding of the quark matter phases in the (T-$\mu_B$)-plane. Strong magnetic fields will likely be also generated at the planned experiments at FAIR, making it possible to explore the region of higher densities under a magnetic field.

The other physical environment where the influence of a magnetic field in the state of quark matter is relevant is the core of neutron stars, which typically are very magnetized objects. From the measured periods and spin down of soft-gamma repeaters (SGR) and anomalous X-ray pulsars (AXP), as well as the observed X-ray luminosities of AXP, certain neutron stars known as magnetars have been found to exhibit surface magnetic fields as large as $10^{14}-10^{16}$ G \cite{Magnetars}. Moreover, since the stellar medium has a very high electric conductivity, the magnetic flux should be conserved. Hence, it is natural to expect a stronger field strength with increasing matter density at the core.  The interior magnetic fields are, however, not directly accessible to observation, thus one can only estimate their values with heuristic methods. Estimates based on macroscopic and microscopic analysis, considering both gravitationally bound and self-bound stars, have led to maximum fields within the range $10^{18}-10^{20}$ G, depending if the inner medium is formed by neutrons \cite{virial}, or quarks \cite{EoS-H}.

A magnetic field is known to induce nontrivial effects in quark matter. In heavy-ion collisions, the high temperature generated during the collisions can induce sphaleron-type transitions to gluon configurations with nonzero winding number. Under these conditions, a strong magnetic field could serve to probe topological nontrivial gluon configurations through the observation of charge separation via the chiral magnetic effect mechanism \cite{CME}. In the other extreme of the QCD phase map, in the region of low temperatures and high densities, a magnetic field can modify the color superconducting phase producing the so-called magnetic color-flavor-locked phase \cite{MCFL}. Furthermore, when the field strength becomes comparable to various characteristic scales -superconducting gap, gluon Meissner mass, and chemical potential- different effects and magnetic phases emerge \cite{Phases}. These effects on dense matter are of interest for astrophysics, as a strong field may affect the matter phase realized in the core and lead to observable signatures through the modification of the equation of state (EoS), transport properties, and others \cite{Review}.

A magnetic field is known to produce the catalysis of chiral symmetry breaking (MC$\chi$SB) \cite{MC} in any system of fermions with arbitrarily weak attractive interaction.  The mechanism responsible for such effect is related to the dimensional reduction of the infrared dynamics of the particles in the lowest Landau level (LLL) \cite{MC}. Such a reduction favors the formation of a chiral condensate because there is no energy gap between the infrared fermions in the LLL and the antiparticles in the Dirac sea.  The MC$\chi$SB modifies the vacuum properties and induces dynamical parameters that depend on the applied field. This effect has been actively investigated for the last two decades \cite{severalaspects}-\cite{prlFIS}. In the original studies of the MC$\chi$SB ~\cite{MC}-\cite{leungwang}, the catalyzed chiral condensate was assumed to generate only a fermion dynamical mass. Recently, however, it has been shown that in QED \cite{ferrerincera}  the
MC$\chi$SB leads to a dynamical fermion mass and inevitably also to a dynamical anomalous magnetic moment (AMM). This is connected to the fact that the AMM does not break any symmetry that has not already been broken by the other condensate. The dynamical AMM in massless QED leads, in turn, to a nonperturbative Lande g-factor and Bohr magneton proportional to the inverse of the dynamical mass. The induction of the AMM also yields a nonperturbative Zeeman effect \cite{ferrerincera}. An important aspect of the MC$\chi$SB is its universal character and hence one expects that the dynamical generation of the AMM should permeate all the models of interacting massless fermions in a magnetic field. Notice that the MC$\chi$SB has been proposed as the mechanism explaining various effects in quasiplanar condensed matter systems~\cite{MC-Applications}, so the additional condensate can be physically relevant for those systems. 

Of particular interest for the present paper is the influence of a magnetic field on the QCD chiral transition. Given that QCD-lattice calculations in the presence of a magnetic field at finite temperature but zero density are feasible, they provide an alternative reliable method to investigate the influence of a magnetic field in the chiral transition in nonperturbative QCD. In this context, a recent result \cite{QCD-latticeB} has shown that while the chiral condensate increases with the applied magnetic field, the critical temperature for chiral symmetry restoration, $T_{C\chi}$, decreases. This result is in contradiction with the fact that the explicit magnetic-field dependence of the dynamical mass obtained through MC$\chi$SB is such that it increases with an increasing field. Therefore, the critical temperature $T_{C\chi}$, which is proportional to the induced dynamical mass at zero temperature, should apparently increase. One possible explanation is that since the MC$\chi$SB is essentially a LLL effect,  the fluctuations produced by a finite temperature will tend to take the quarks out of the LLL and hence cancel the magnetic catalysis effect, unless the field is much larger than T. However, strictly speaking, once the system is in the region of supercritical coupling, the effect of the magnetic field is not exactly described by the MC$\chi$SB phenomenon, since now a nonzero constituent mass is generated even at zero magnetic field.  The magnetic field notwithstanding increases the value of this dynamical mass at zero T, and one would expect that this would lead to a higher critical temperature, in contradiction with the lattice results.  Although certain attempts to explain those contradictory findings already exist in the literature \cite{Kojo}, it is still an open question under scrutiny.

In the present paper we investigate the dynamical generation of a net magnetic moment in the ground state of a one-flavor Nambu-Jona-Lasinio (NJL) model in a magnetic field and discuss its implications for the chiral phase transition at finite temperature. Notice that the chiral condensate is formed from the pairing of quarks and antiquarks with opposite spins. The dynamical mass induced by the chiral condensate embeds each quark and antiquark with an AMM. The AMMs of the quarks/antiquarks in the pair point in the same direction, so the pair has a nonzero magnetic moment (MM). Then, the presence of a magnetic field breaks the Lorentz symmetry and, as in the case of massive particles \cite{Ioffe}, it allows the generation of a nonzero vacuum expectation value for $<0|\overline{q}\sigma^{\mu\nu}q|0>$, where $q$ is the quark field.  Such a vev accounts for the net magnetic moment of the ground state.  The two condensates contribute to the effective dynamical mass resulting in a significant increase in the critical temperature for the chiral restoration, as compared to the case where only the magnetically catalyzed chiral condensate is considered.

The paper is organized as follows. In Sec.II, we introduce the one-flavor NJL model, which includes four-fermion interactions consistent with the Fierz identities in the presence of a uniform magnetic field. The corresponding mean-field effective potential in the presence of the magnetic field is then calculated in Sec. III and used in Sec. IV to obtain the condensate solutions in the presence of a magnetic field. In Sec. V, the critical temperature for chiral symmetry restoration is calculated. In Sec. VI we summarize the paper results, its implications, and comment on future works. In the Appendix, we detail the Fierz transformations for particle/antiparticle channels in a system with broken rotational symmetry.

\section{NJL Model in a magnetic field}

Our main goal here is to investigate the effect of a constant and homogeneous magnetic field in the spontaneous breaking of chiral symmetry in QCD.  With this goal, we are going to use a simple NJL model that can be interpreted as the result of integrating out the gluon fields and quark fluctuations with momenta larger than some scale $\Lambda$, with $\Lambda \gtrsim \Lambda_{QCD}$. Our NJL model has four-fermion point interactions that capture several ingredients of QCD chiral symmetry in a magnetic field, but fails to describe the phenomenon of confinement. The use of NJL models to explore chiral symmetry breaking in QCD with nonzero magnetic field has been a successful strategy followed by many previous works \cite{CSB}. The new element in the present investigation will be the introduction of a yet unexplored four-fermion channel that becomes relevant only in the presence of a magnetic field and can lead to nontrivial physical consequences.

With the above goal in mind, let us consider the following  NJL model of massless quarks in the presence of a constant and uniform magnetic field
\begin{equation}\label{lagrangian}
\mathcal{L}=\bar{\psi}i\gamma^{\mu}D_{\mu}\psi+\mathcal{L}_{int}^{(1)}+\mathcal{L}_{int}^{(2)}
\end{equation}
The single-flavor Dirac spinor $\psi$  belongs to the fundamental representation of the SU(N$_c$) color group. The electromagnetic four-potential in the covariant derivative $D_\mu=\partial_\mu+iqA^{ext}_\mu$ can be chosen, without loss of generality, in the gauge $A^{(ext)}_\mu=(0,0,Bx_1,0)$,  so to have a constant and homogenous magnetic field of magnitude $B$ pointing in the $x_3$-direction. We use, from now on, the Lorentz metric $\eta_{\mu\nu}=(1,-\overrightarrow{1})$ and the Dirac matrices in the chiral representation. The interaction
\begin{equation}\label{lagrangian-1}
\mathcal{L}_{int}^{(1)}=\frac{G}{2}[(\bar{\psi}\psi)^2+(\bar{\psi}i\gamma^5\psi)^2],
\end{equation}
has the conventional four-fermion scalar and pseudoscalar channels used in many previous studies based on NJL models \cite{CSB}.
In addition, we introduce a new channel
\begin{equation}\label{lagrangian-2}
\mathcal{L}_{int}^{(2)}=\frac{G'}{2}[(\bar{\psi}\Sigma^3\psi)^2+(\bar{\psi}i\gamma^5\Sigma^3\psi)^2],
\end{equation}
that preserves chiral symmetry and rotations about the magnetic field direction. Here $\Sigma^3=\frac{i}{2}[\gamma^1, \gamma^2]= \sigma_{\bot}^{\mu \nu}$ is the spin operator in the direction of the applied field. In (\ref{lagrangian})-(\ref{lagrangian-2}), summation over color index has been assumed.

The new interaction channel $\mathcal{L}_{int}^{(2)}$ with second-rank tensor structure naturally emerges using the Fierz identities in the one-gluon-exchange channels of QCD when the rotational symmetry is broken. To understand this, one recalls that a magnetic field always selects a preferable direction and explicitly breaks the rotational symmetry, reducing it to the subgroup $O(2)$ of spatial rotations about the field direction. This in turn implies that the tensor structures of the Dirac ring split in components parallel and transverse to the field direction with the help of the normalized tensor  $\widehat{F}_{\mu\nu}=F_{\mu\nu}/|B|$,
\begin{equation}
\gamma^\|=\eta_{\|}^{\mu \nu}\gamma_\nu, \quad \gamma^\bot=\eta_{\perp}^{\mu \nu}\gamma_\nu
\label{gammas}
\end{equation}
with
\begin{equation}
\eta_{\|}^{\mu \nu}=\eta^{\mu \nu}-\widehat{F}^{\mu \rho}\widehat{F}_{\rho}^\nu, \quad \eta_{\perp}^{\mu \nu}=\widehat{F}^{\mu \rho}\widehat{F}_{\rho}^\nu.
\label{etas}
\end{equation}
being the longitudinal and transverse Minkowskian metric tensors respectively. In the rest frame, for a magnetic field in the $x_3$ direction, $\eta_{\|}^{\mu \nu}$ has only $\mu, \nu = 0,3$ components, and $\eta_{\bot}^{\mu \nu}$ has $\mu, \nu =1,2$.

As a consequence, the four-fermion interaction Lagrangian density separates in two terms,
\begin{equation}
\mathcal{L}_{int}=\frac{g_\|^2}{2 \Lambda^2}(\bar{\psi}\gamma_{\|}^{\mu}\psi)(\bar{\psi}\gamma^{\|}_\mu \psi)+\frac{g_\bot^2}{2\Lambda^2}(\bar{\psi}\gamma_{\bot}^\mu \psi)(\bar{\psi}\gamma^{\bot}_\mu \psi).
\label{Int-Lag}
\end{equation}
Notice that despite the fact that there is no direct coupling between the gluons and the magnetic field, the vertex with the fermions is modified because of the distinction between longitudinal and transverse fermion modes in this case. The extreme case occurs for magnetic fields of the order of the energy scale of the fermions, where all the fermions are in the LLL and hence the only modes entering in the bare coupling are the longitudinal ones. This is the origin of the anisotropy in the strong-coupling vertex in the presence of a magnetic field. Therefore, this anisotropy should be reflected in the NJL model in a magnetic field.

 On the other hand, as detailed  in the Appendix, the $O(3)\rightarrow O(2)$ symmetry breaking that takes place in the presence of a magnetic field leads to the anisotropic Fierz identities
\begin{equation}\label{Fierz-Ident-paralle}
( \gamma_{\|}^\mu)_{il} \left ( \gamma^{\|}_\mu \right )_{kj}=\frac{1}{2}\left \{  \left ( 1 \right )_{il}\left ( 1 \right )_{kj}+(i\gamma_5)_{il}(i\gamma_5)_{kj}
+\frac{1}{2}\left ( \sigma_{\bot}^{\mu\nu} \right )_{il}\left ( \sigma^{\bot}_{\mu\nu} \right )_{kj}-\left ( \sigma^{03} \right )_{il}\left ( \sigma_{03} \right )_{kj}+...
 \right\},
\end{equation}
and
\begin{equation}\label{Fierz-Ident-perp}
( \gamma_{\bot}^\mu )_{il} \left (\gamma^{\bot}_\mu \right )_{kj}=\frac{1}{2}\left \{  \left ( 1 \right )_{il}\left ( 1 \right )_{kj}+(i\gamma_5)_{il}(i\gamma_5)_{kj}
-\frac{1}{2}\left ( \sigma_{\bot}^{\mu\nu} \right )_{il}\left (\sigma^{\bot}_{\mu\nu} \right )_{kj}+\left ( \sigma^{03} \right )_{il}\left ( \sigma_{03} \right )_{kj}+...
 \right\},
\end{equation}
where $\|$ and $\bot$ denotes parallel $\mu=(0,3)$ and transverse $\mu=(1,2)$ Lorentz indexes with respect to the magnetic field direction. Einstein summation convention for repeated indices is assumed.

From (\ref{lagrangian-1}), (\ref{lagrangian-2}), (\ref{Int-Lag})-(\ref{Fierz-Ident-perp}), one can readily identify the channels considered in $\mathcal{L}_{int}^{(1)}$ and $\mathcal{L}_{int}^{(2)}$. Then, the couplings $G$ and $G'$ can be related to $g_{\|}$ and $g_{\bot}$ through
\begin{equation}\label{G-couplingsn}
G=(g_\|^2+g_\bot^2)/2\Lambda^2, \qquad G'=(g^2_\|-g^2_\bot)/2\Lambda^2
\end{equation}
with $\Lambda$ the energy scale of the effective theory. At zero magnetic field  $g=g_\|=g_\bot$ and one can use measured physical quantities to find consistent values for $G$ and $\Lambda$. However, at nonzero magnetic field there are no measured parameters that can be used for this purpose. In lieu of arbitrarily assigning values to $G$, $G'$ and $\Lambda$, we can take $G$ and $\Lambda$ at their zero-field values, chosen to fix the pion decay constant to $f_\pi = 93$ MeV and the condensate density per quark  to $<\overline{u}u>=-(250  \textrm{MeV})^3$, and then assign values to $G'$ with the constraint $G'\leq G$. Notice that $G' \geq 0$ because when the field increases, so does the occupation of the LLL, hence reinforcing the longitudinal contributions over the transverse ones.

The Lagrangian density (\ref{lagrangian}) can be also interpreted as an ad-hoc single-flavor effective theory consistent with the symmetries of QCD in a magnetic field. Apart from the subgroup of rotations already mentioned, it is also invariant under baryon symmetry, $U(1)_B$, and because of the absence in (\ref{lagrangian}) of a fermion mass, chiral symmetry $U(1)_\chi$ is preserved. For other contexts where unconventional four-point interactions in NJL-like models have been considered see \cite{AMM-CS, Klimt, Kitasawaetal, NJL-Quarky}.

\section{Effective Potential in the mean-field approximation}
Let us explore now the possibility of the following homogeneous condensates
\begin{equation}\label{condensates}
\langle\bar{\psi}\psi\rangle=-\frac{\sigma}{G}, \qquad \langle\bar{\psi}i\gamma^5\psi\rangle=-\frac{\Pi}{G}, \qquad  \langle\bar{\psi}i\gamma^1 \gamma^2\psi\rangle=-\frac{\xi}{G'}, \qquad  \langle\bar{\psi}i\gamma^0 \gamma^3\psi\rangle=-\frac{\xi'}{G'},
\end{equation}
where $\sigma$, $\Pi$, $\xi$ and $\xi'$ are constant parameters.

Using them to perform the Hubbard-Stratanovich transformation in the Lagrangian density (\ref{lagrangian}), we obtain the partition function in the mean-field approximation
\begin{equation}\label{Z}
Z=\int D[\overline{\psi}]D[\psi]exp\left(iS(\sigma,\Pi,\xi,\xi')\right),
\end{equation}
with action
\begin{equation}\label{lagrangian-cond}
S(\sigma,\Pi,\xi,\xi')=\int d^4x \bar{\psi}(x)(i\gamma^{\mu}D_{\mu}-\sigma-i\gamma^5 \Pi-i\gamma^1 \gamma^2 \xi-i\gamma^0 \gamma^3 \xi')\psi(x)-\frac{V}{2G}(\sigma^2+\Pi^2)-\frac{V}{2G'}(\xi^2+\xi'^2).
\end{equation}

The corresponding mean-field effective potential is
\begin{equation}\label{effect-pot2}
\Omega(\sigma,\Pi,\xi,\xi')=\frac{\sigma^2+\Pi^2}{2G}+\frac{\xi^2+\xi'^2}{2G'}+\frac{i}{V}\textrm{Tr}\textrm{ln}(iD\cdot\gamma-\sigma
-i\gamma^5\Pi-i\gamma^1\gamma^2\xi-i\gamma^0\gamma^3\xi')
\end{equation}
where the trace (Tr) acts in color, Dirac, and coordinate spaces.

At this point, it is convenient to transform to momentum space with the help of the Ritus transformation \cite{Ritus:1978cj}. This method is based on a Fourier-like transformation that uses eigenfunction matrices $E_p(x)$. The $E_p(x)$ are the wave functions of the asymptotic states of charged
fermions in a uniform magnetic field. The method yields a fermion Green function that is diagonal in momentum space and explicitly dependent on the Landau levels. Although valid at any field strength, this formalism is particularly convenient to study the strong-field region, where the main contribution comes from the LLL \cite{leungwang, ferrerincera, MCFL}.

Using Ritus's approach, the inverse propagator in momentum space \cite{ferrerincera} takes the form
\begin{eqnarray}\label{Inv-Propagator}
G^{-1}_{l}(p,p')= \int d^4x d^4x'
\overline{E}_{p}^{l}(x)[iD\cdot\gamma-\sigma-i\gamma^5\Pi-i\gamma^1\gamma^2\xi-i\gamma^0\gamma^3\xi')]\delta^{(4)}(x-x')E_{p'}^{l'}(x')=\nonumber
 \\
=(2\pi)^4\widehat{\delta}^{(4)}(p-p')\Theta(l)\widetilde{G}^{-1}_l(\overline{p})\quad \quad \quad \quad \quad \quad
\end{eqnarray}
with
\begin{equation}
\widetilde{G}^{-1}_l(\overline{p})=[{\overline{p}}\cdot\gamma
-\sigma-i\gamma^5\Pi-i\gamma^1\gamma^2\xi-i\gamma^0\gamma^3\xi'] \label{Inv-Propagator-Def},
\end{equation}
and
\begin{equation}
 \overline{p}^\mu=(p^{0},0, -\textrm{sgn}(qB)\sqrt{2|qB| l},p^{3}).
 \end{equation}

The $E_{p}^{l}(x)$ are matrix functions given as a linear combination of spin up ($+$) and down ($-$) projectors $\Delta(\pm)$. For $q>0$, they can be written as
\begin{equation}\label{Ep}
 E_{p}^{l}(x)=E_{p}^{+}(x)\Delta(+)+E_{p}^{-}(x)\Delta(-),
\end{equation}
with
 \begin{equation}\label{spinproject}
 \Delta(\pm)=\frac{I\pm i\gamma^{1}\gamma^{2}}{2} \quad\quad \textrm{for} \quad q>0,
  \end{equation}
and
\begin{eqnarray}\label{E-x}
E_{p}^{+}(x)=N_{l}e^{-i(p_{0}x^{0}+p_{2}x^{2}+p_{3}x^{3})}D_{l}(\rho),\qquad
\nonumber
\\
E_{p}^{-}(x)=N_{l-1}e^{-i(p_{0}x^{0}+p_{2}x^{2}+p_{3}x^{3})}D_{l-1}(\rho).
\end{eqnarray}
Index $l=0,1,2,...$ is the Landau level number that characterizes the discretization of the transverse momentum in a magnetic field. Here $N_{l}=(4\pi qB)^{1/4}/\sqrt{l!}$ is a normalization constant and
$D_{l}(\rho)$ denotes the parabolic cylinder function of argument
$\rho=\sqrt{2qB}(x_{1}-p_{2}/qB)$ and index $l$.

The coefficient
  \begin{equation}
 \Theta(l)=\Delta(+)\delta^{l0}+I(1-\delta^{l0})
  \end{equation}
 in (\ref{Inv-Propagator}) takes into account the lack of spin degeneracy of the LLL.

To obtain (\ref{Inv-Propagator}) we used the orthogonality of the $E_p^l$ functions
\cite{leungwang}
\begin{equation}
\int d^{4}x \overline{E}_{p}^{l}(x)E_{p'}^{l'}(x)=(2\pi)^4
\widehat{\delta}^{(4)}(p-p')\Theta(l) \ , \label{orthogonality}
\end{equation}
with $\overline{E}_{p}^{l}\equiv \gamma^{0}
(E_{p}^{l})^{\dag}\gamma^{0}$ and $\widehat{\delta}^{(4)}(p-p')=\delta^{ll'} \delta(p_{0}-p'_{0})
\delta(p_{2}-p'_{2}) \delta(p_{3}-p'_{3})$.

After going to Euclidean variables, we can use the completeness relation
\begin{equation}
\sum \hspace{-0.49cm}\int \frac{d^{4}p^E}{\left( 2\pi
\right) ^{4}}E_{p}^{l}(x)  \overline{E}_{p}^{l}(x)= \left( 2\pi
\right) ^{4}
\delta^{(4)}(x-x'), \label{orthogonality}
\end{equation}
to invert (\ref{Inv-Propagator}) and find
\begin{eqnarray}\label{Propagator}
G^{-1}(x,x')= \sum \hspace{-0.49cm}\int \frac{d^{4}p^E}{\left( 2\pi
\right) ^{4}}\sum \hspace{-0.49cm}\int \frac{d^{4}p'^E}{\left( 2\pi
\right) ^{4}} E_{p}^{l}(x) G^{-1}(p,p')\overline{E}_{p'}^{l'}(x').
\end{eqnarray}
with $\sum_{\it l}\hspace{-0.47cm}\int \frac{d^{4}p^E}{\left( 2\pi
\right) ^{4}}\equiv \sum_{l=0}^\infty \int \frac{dp_4dp_2dp_3}{(2\pi)^4}$.

With the help of (\ref{Propagator}), the effective potential can be written as
\begin{eqnarray}\label{Grand-Potential-2}
\Omega(\sigma,\Pi,\xi,\xi')=\frac{\sigma^2+\Pi^2}{2G}+\frac{\xi^2+\xi'^2}{2G'}-N_c qB \textrm{tr} \sum_{l=0}^{\infty}\int_{-\infty}^{\infty}\frac{dp_4dp_3}{(2\pi)^3} \textrm{ln} \Theta(l)\widetilde{G}^{-1}_l(\overline{p})
\end{eqnarray}
 where the integration in $p_2$ was done using
\begin{equation}
\int_{-\infty}^\infty\frac{dp_2}{2\pi}=\int_{-\infty}^{\infty}\frac{dp_2}{2\pi}  e^{-i\frac{p_2p_1}{qB}} |_{p_1=0} =
\frac{1}{l_B^2}\delta(p_1)|_{p_1=0} =\frac{1}{l_B^2}\int_{-\infty}^{\infty} dx_1   \label{Int-p2},
\end{equation}
and the trace (tr) now only acts on the spinorial matrices. Here $l_B=1/\sqrt{qB}$ denotes the magnetic length.

Taking into account that the $l=0$ term only gets contributions from the subspace of spinors with a single spin projection; spin up (down) for $q>0$  ($q<0$); it can be separated from the rest to write
\begin{equation}\label{Grand-Potential-4}
\Omega(\sigma,\Pi,\xi,\xi')= \frac{\sigma^2+\Pi^2}{2G}+\frac{\xi^2+\xi'^2}{2G'}-N_c qB\left[\int_{-\infty}^{\infty}\frac{dp_4dp_3}{(2\pi)^3} \ln \det \widetilde{G}^{-1}_0(\overline{p})+ \sum_{l=1}^{\infty}\int_{-\infty}^{\infty}\frac{dp_4dp_3}{(2\pi)^3} \ln \det \widetilde{G}^{-1}_l(\overline{p})\right]
\end{equation}

Integrating in $p_4$ we find
\begin{equation}\label{Effective-Potential-LLL-2}
\Omega(\sigma,\Pi,\xi,\xi')= \frac{\sigma^2+\Pi^2}{2G}+\frac{\xi^2+\xi'^2}{2G'}-\frac{N_c qB}{4\pi^2}\int_{-\infty}^{\infty }|\varepsilon_0 |dp_3-\frac{N_c qB}{4\pi^2}\sum_{\eta=\pm 1}\sum_{l=1}^{\infty}\int_{-\infty }^{\infty }|\varepsilon_{l,\eta} |dp_3,
\end{equation}

with energy spectrum
\begin{eqnarray}\label{Spectrum}
\varepsilon_0^2=p_3^2+(\sigma+\xi)^2+(\Pi+\xi')^2, \quad l=0, \qquad \qquad \qquad \qquad \qquad \qquad
\nonumber
\\
\varepsilon_{l,\eta}^2=p_3^2+\Pi^2+\xi'^2+\sigma^2(1-X)+2lqB(1-X')+\left(\sqrt{\sigma^2X+2lqB}+\eta \xi \right)^2, \quad l\geq 1, \quad \eta=\pm 1
\end{eqnarray}

where
\begin{equation}\label{Relations}
X=\left (1+\frac{\Pi}{\sigma}\frac{\xi'}{\xi}\right)^2, \qquad X'=\left (1+\frac{\xi'^2}{\xi^2}\right)
\end{equation}

The factor $qB/4\pi^2$  accounts for the density of states of the Landau levels. The spectrum of the quasiparticles with Landau levels $l\geq 1$ exhibits a Zeeman splitting ($\eta=\pm1$) indicating that the new dynamical parameter $\xi$ enters as an AMM energy term. This is even more evident if we take $\Pi=\xi'=0$ in the spectrum, since it becomes equal to the one found in QED with dynamical mass and AMM \cite{ferrerincera}. No splitting is present in the $l=0$ mode, in agreement with the fact that the fermions in the LLL only has one spin projection.

\section{Condensate Solutions}

\subsection{Gap Equations}

We are interested in the situation where the magnetic field is large enough to have all the quarks lying in the LLL, thus the ground state is dominated by the infrared dynamics and only the first integral in the RHS of (\ref{Effective-Potential-LLL-2}) contributes to the equations. This requires magnetic fields $qB \sim \Lambda^2 \gtrsim\Lambda^2_{QCD}$. Such large fields are actually generated in off-central heavy-ion collisions at RHIC.

To determine the dynamical solutions for the four condensates $\sigma$, $\Pi$, $\xi$ and $\xi'$, we need to solve the gap equations
\begin{eqnarray}\label{Gap-Eqs}
\frac{\partial\Omega(\sigma,\Pi,\xi,\xi')}{\partial\sigma}= \frac{\sigma}{G}-(\sigma+\xi){\cal{I}}_0=0, \qquad \frac{\partial\Omega(\sigma,\Pi,\xi,\xi')}{\partial\xi}= \frac{\xi}{G'}-(\sigma+\xi){\cal{I}}_0=0, \nonumber
 \\
 \frac{\partial\Omega(\sigma,\Pi,\xi,\xi')}{\partial\Pi}= \frac{\Pi}{G}-(\Pi+\xi'){\cal{I}}_0=0, \qquad
 \frac{\partial\Omega(\sigma,\Pi,\xi,\xi')}{\partial\xi'}= \frac{\xi'}{G'}-(\Pi+\xi'){\cal{I}}_0=0,
\end{eqnarray}
where
\begin{equation}\label{Integral}
{\cal{I}}_0=\frac{N_cqB}{2\pi^2}\int_{0}^{\Lambda}\frac{dp_3}{\varepsilon_0}
\end{equation}
Here we introduced the momentum cutoff $\Lambda$ below which the NJL theory is valid.
One can check that the solution of (\ref{Gap-Eqs}) satisfies
\begin{equation}\label{Cond-Sol-1}
\overline{\xi}=\frac{G'}{G}\overline{\sigma}, \qquad  \overline{\xi'}=\frac{G'}{G}\overline{\Pi}
\end{equation}
Then, the condensates can be found from
\begin{equation}\label{Cond-Sol-2}
\int_0^\Lambda \frac{dp_3}{\sqrt{p_3^2+(1+\frac{G'}{G})^2(\overline{\sigma}^2+\overline{\Pi}^2)}}=\frac{2\pi^2}{(G+G')N_cqB}
\end{equation}
Notice that the gap equation (\ref{Cond-Sol-2}) depends only on the $U_L(1)\times U_R(1)$-invariant $\overline{\sigma}^2+\overline{\Pi}^2$, a typical feature of the MC$\chi$SB phenomenon \cite{CSB,MC}. Hence, we can, as usual, specialize the condensate configuration with $\Pi=0$ and $\sigma$ constant. As expected for a magnetically catalyzed condensate, no critical coupling is needed for a nontrivial solution to exist.

From (\ref{Cond-Sol-1}), we see that no solution exists with $\overline{\sigma}\neq 0$ and $\overline{\xi}=0$, and vice versa. The energetically favored solution has expectation values of both $\overline{\sigma}$ and $\overline{\xi}$ different from zero. In the same way that the chiral condensate $\langle \overline{\psi}\psi  \rangle$ gives a dynamical mass to the quasiparticles, the new condensate $\langle \overline{\psi}i\gamma^1 \gamma^2\psi\rangle$ gives them a dynamical AMM. Once the quarks acquire a dynamical mass, they should also acquire a dynamical AMM. This effect has been found to occur in QED \cite{ferrerincera} and the appearance of the condensate $\xi$ in our NJL model is a clear indication that it also occurs in QCD.  One can understand the inevitability of a dynamical AMM in the magnetically catalyzed system on the base of symmetry arguments. Once the chiral symmetry is dynamically broken, there is no symmetry protection for the AMM, because it breaks the exact same symmetry. The AMM of the quarks leads to a nonzero dynamical MM for the pair. That the pairs should have a dynamical MM is easy to understand, since they are formed by quarks and antiquarks with opposite spins, so the fermions' AMMs point in the same direction. The magnetic field aligns the pairs's MM leading to a net MM of the ground state.

\subsection{Effect on the Quasiparticle's Effective Mass}

The solutions of the gap equations (\ref{Cond-Sol-1})-(\ref{Cond-Sol-2}) are
\begin{equation}\label{sigma}
\overline{\sigma}=\left(\frac{2G\Lambda}{G+G'}\right )\exp{-\left[\frac{2\pi^2}{(G+G')N_cqB}\right]}
\end{equation}
and
\begin{equation}\label{AMM}
\overline{\xi}=\left(\frac{2G'\Lambda}{G+G'}\right )\exp{-\left[\frac{2\pi^2}{(G+G')N_cqB}\right]}
\end{equation}

It is worthwhile to underline that the induced AMM term (\ref{AMM}) depends nonperturbatively on the coupling constant and the magnetic field. This behavior reflects two important facts: (i) in a massless theory, chiral symmetry can be only broken dynamically, that is, nonperturbatively; and (ii) the MC$\chi$SB phenomenon is essentially a LLL effect. The LLL plays a special role due to the absence of a gap between it and the Dirac sea. The rest of the LLs are separated from the Dirac sea by energy gaps that are multiples of $\sqrt{2qB}$, and hence do not significantly participate in the pairing mechanism at the subcritical couplings where the magnetic catalysis phenomenon is relevant. Since the dynamical generation of the AMM is produced mainly by the LLL pairing dynamics, one should not expect to obtain a linear-in-B AMM term, even at weak fields, in sharp contrast with the AMM appearing in theories of massive fermions. In the latter case, not only the AMM is obtained perturbatively through radiative corrections, but considering the weak-field approximation means first summing in all the LL's, which contribute on the same footing, and then taking the leading term in an expansion in powers of B \cite{Ioffe, Schwinger}.  Notice that such a linear dependence does not hold, even in the massive case, if the field is strong enough to put all the fermions in the LLL \cite{Jancovici}.

The effect of the new condensate $\langle\bar{\psi}i\gamma^1 \gamma^2\psi\rangle$ is to increase the effective dynamical mass of the quasiparticles in the LLL,
 \begin{equation}\label{neweffectivemass}
M_{\xi}=\overline{\sigma}+\overline{\xi}=2\Lambda\exp{-\left[\frac{2\pi^2}{(G+G')N_cqB}\right]}
\end{equation}

In QCD, for fields, $qB \sim \Lambda^2$, the dimensional reduction of the LLL fermions would constraint the LLL quarks to couple with the gluons only through the longitudinal components. Thus, to consistently work in this regime within the NJL model, we should consider, taking into account (\ref{G-couplingsn}), that $G' = G$, so that $G+G'=2G$.

Because the effective coupling enters in the exponential, the modification of the dynamical mass by the magnetic moment condensate can be significant. As a consequence, the quasiparticles should be much heavier in our model than in previous studies that ignored the magnetic moment interaction \cite{CSB}. How much heavier can be estimated from the logarithm of the ratio between the effective mass (\ref{neweffectivemass}) and the mass found with G' equal to zero (i.e. $M_{\xi=0}=2\Lambda\exp{-[2\pi^2/GN_cqB]}$)

\begin{equation}\label{rest-energy-ratio}
\ln \left (\frac{M_\xi}{M_{\xi=0}}\right )=\frac{2\pi^2}{GN_cqB}\left(\frac{\eta}{1+\eta}\right)
\end{equation}
Here we used $G'=\eta G$, but we know that for $qB/\Lambda^2\sim 1$, $\eta \simeq 1$. Using the values $G\Lambda^2=1.835$, $\Lambda=602.3$ MeV \cite{G-Interaction}, $N_c=3$ and $q=|e|/3\simeq 0.1$,  we estimate the RHS of (\ref{rest-energy-ratio}) as $(\pi^2/GN_cqB)\simeq 1.8$. Due to the condensate $\xi$ the dynamical mass of the quasiparticles increases sixfold. This result shows that at strong fields the new channel of interaction must not be ignored, as it may lead to important physical consequences. One of them is the critical temperature for the chiral restoration, as we show in the next section.

\section{Critical temperature}

\subsection{Condensate solutions at finite temperature}

Our goal now is to calculate the critical temperature for chiral symmetry restoration in the magnetized system. With that aim, we take the LLL approximation in  \ref{Grand-Potential-4} and replace the integration in $p_4$ by  the Matsubara's sum 
 \begin{equation}\label{Matsubara}
\int_{-\Lambda}^\Lambda\frac{dp_4}{2\pi}\rightarrow\frac{1}{\beta}\sum_{p_4},\qquad  \beta= \frac{1}{T}, \qquad p_4=\frac{(2n+1)\pi}{\beta}, \quad n=0,\pm1,\pm2,...
\end{equation}
to obtain
\begin{equation}\label{Thermo-Potential}
\Omega_0^T(\sigma,\xi)=\frac{\sigma^2+\Pi^2}{2G}+\frac{\xi^2+\xi'^2}{2G'}-\frac{N_cqB}{\beta}\int_{-\infty}^\infty \frac{dp_3}{4\pi^2}\sum_{p_4}\ln \left[p_4^2+\varepsilon_0^2\right] 
\end{equation} 
Performing the sum in $p_4$ \cite{Kapusta} and introducing the momentum cutoff $\Lambda$, we obtain
\begin{equation}\label{Thermo-Potential-2}
\Omega_0^T(\sigma,\xi)=-N_cqB\int_{0}^\Lambda \frac{dp_3}{2\pi^2}\left[ \varepsilon_0 +\frac{2}{\beta}\ln \left(1+e^{-\beta\varepsilon_0}\right)\right]+\frac{\sigma^2+\Pi^2}{2G}+\frac{\xi^2+\xi'^2}{2G'}
\end{equation}

The gap equations at finite temperature are then given by
\begin{eqnarray}\label{Gap-Eqs-T}
\frac{\partial\Omega_0^T(\sigma,\Pi,\xi,\xi')}{\partial\sigma}= \frac{\sigma}{G}-(\sigma+\xi)[{\cal{I}}_0+{\cal{I}}_\beta]=0, \qquad \frac{\partial\Omega_0^T(\sigma,\Pi,\xi,\xi')}{\partial\xi}= \frac{\xi}{G'}-(\sigma+\xi)[{\cal{I}}_0+{\cal{I}}_\beta]=0, \nonumber
 \\
 \frac{\partial\Omega_0^T(\sigma,\Pi,\xi,\xi')}{\partial\Pi}= \frac{\Pi}{G}-(\Pi+\xi')[{\cal{I}}_0+{\cal{I}}_\beta]=0, \qquad
 \frac{\partial\Omega_0^T(\sigma,\Pi,\xi,\xi')}{\partial\xi'}= \frac{\xi'}{G'}-(\Pi+\xi')[{\cal{I}}_0+{\cal{I}}_\beta]=0,
\end{eqnarray}
where ${\cal{I}}_0$ is defined in (\ref{Integral}), and
\begin{equation}\label{Integral-T}
{\cal{I}}_\beta=\frac{N_cqB}{2\pi^2}\int_{0}^{\Lambda}\frac{dp_3}{\varepsilon_0}\frac{2e^{-\beta\varepsilon_0/2}}{e^{\beta\varepsilon_0/2}+e^{-\beta\varepsilon_0/2}}
\end{equation}
\begin{figure} \label{Fig1}
\begin{center}
\includegraphics[width=0.6\textwidth]{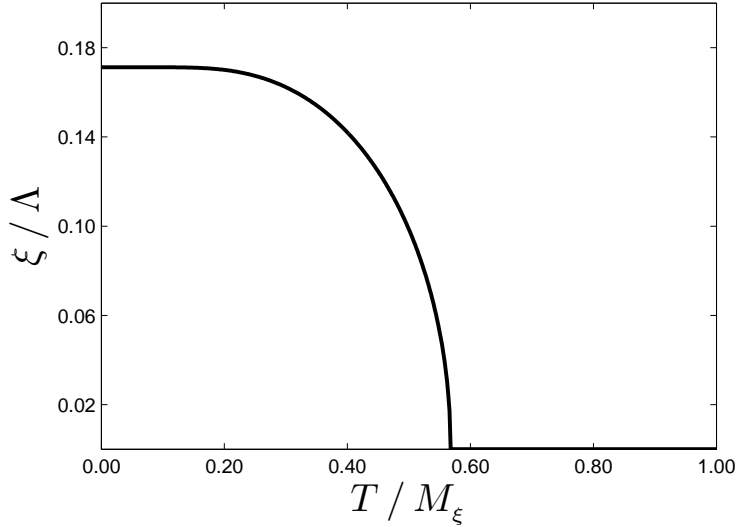}
\caption{\footnotesize Condensate $\xi$, normalized by $\Lambda$, as a function of the temperature, $T$, normalized by the zero-temperature dynamical mass $M_\xi$.} \label{TcX}
\end{center}
\end{figure}
Once again we can check that the solutions of (\ref{Gap-Eqs-T}) satisfy relations similar to those in Eq. (\ref{Cond-Sol-1}).
Then, the condensates can be found from
\begin{equation}\label{Cond-Sol-2-T}
\int_0^\Lambda \frac{dp_3}{\sqrt{p_3^2+(1+\frac{G}{G'})^2(\overline{\xi}^2+\overline{\xi'}^2)}}\tanh \left(\frac{\sqrt{p_3^2+(1+\frac{G}{G'})^2(\overline{\xi}^2+\overline{\xi'}^2)}}{2T}\right)=\frac{2\pi^2}{(G+G')N_cqB}
\end{equation}

Just as in vacuum, the gap equation (\ref{Cond-Sol-2-T}) depends only on the $U_L(1)\times U_R(1)$-invariant $\overline{\xi}^2+\overline{\xi'}^2$. Hence, we can, as usual, specialize the condensate configuration along the $\xi$ internal direction and take $\xi'=0$. In Fig.1 we represent the numerical solution of (\ref{Cond-Sol-2-T}). Notice that the condensate $\xi$ decreases continuously with the temperature, vanishing at $T \sim 0.6 M_\xi$, with $M_\xi$ the zero-T dynamical mass. This behavior is consistent with the order parameter of a second-order transition. Equally important, the chiral condensate $\overline{\sigma}$ evaporates together with $\overline{\xi}$, because of the relations (\ref{Cond-Sol-1}), which, as already pointed out, remain valid at finite temperature.

\subsection{Critical-temperature analytical expression} 

The critical temperature $T_{C_\chi}$ can be analytically found from the condition
\begin{equation}\label{Critical-Temperature}
\frac{\partial^2\Omega_0^{T_{C_\chi}}}{\partial \overline{\sigma}^2}|_{\overline{\sigma}=\overline{\xi}=0}=\frac{\sigma^2+\Pi^2}{2G}+\frac{\xi^2+\xi'^2}{2G'}-\frac{N_cqB}{2\pi^2}\left [\frac{G+G'}{G}\int_{0}^\Lambda \frac{dp_3}{p_3}\tanh \left(\frac{\beta_{C_\chi} p_3}{2} \right ) +\frac{2\pi^2}{GN_cqB}\right]=0
\end{equation}
We would have arrived at the same condition by taking instead the derivative with respect to $\overline{\xi}$. This is a consequence of the proportionality between $\overline{\sigma}$ and $\overline{\xi}$, given in Eq. (\ref{Cond-Sol-1}), which implies that  the two condensates evaporate at the same critical temperature.

Doing the change $p_3 \rightarrow p_3/T_{C_\chi}$, we have
\begin{equation}\label{Critical-Temperature-2}
\int_{0}^{\Lambda/T_{C_\chi}} \frac{dp_3}{p_3}\tanh \left (\frac{p_3}{2} \right ) =\frac{2\pi^2}{(G+G')N_cqB},
\end{equation}
so the resulting critical temperature is
\begin{equation}\label{Critical-Temperature-3}
T_{C_\chi}=1.16\Lambda \exp -\left [\frac{2\pi^2}{(G+G')N_cqB}\right ]=0.58M_\xi
\end{equation}
in agreement with the result found numerically in Fig. 1. The fact that the critical temperature is proportional to the dynamical mass at zero temperature, is consistent with the findings in other models \cite{Temperature}. In the present case, since the dynamical mass is increased by the AMM, the critical temperature is proportionally increased.

That the chiral transition is second order can be seen directly from Fig.1, as well as analytically, from the positiveness of the second derivative of $\Omega$ near the phase transition,
\begin{equation}\label{Critical-Temperature-4}
\frac{\partial^2\Omega}{\partial(\bar{\sigma}^2)^2}|_{\beta\approx\beta_C}\approx\frac{N_c qB\beta}{32\pi^2} \left[\frac{G+G'}{G}\right]^4\int_0^{\beta\Lambda/2}dz \frac{\tanh z}{z^3}\left[1-\frac{z}{\sinh z \cosh z} \right]>0.
\end{equation}

We underline that the existence of a unique critical temperature for the evaporation of the two condensates reflects the fact that the condensate $\xi$ does not break any new symmetry that was not already broken by the condensate $\sigma$ and the magnetic field, as discussed above. The simultaneous evaporation of the chiral and magnetic moment condensates has been also reported in the context of lattice QCD \cite{Costa}. There are, however, important differences in the way the magnetic field influences the system in lattice QCD and in the situation considered in the present work. In Ref. \cite{Costa} the coupling is supercritical, so the quark have constituent masses even at zero field and the tensor term can be considered to be linear in B. In our case, however, the quarks acquire their mass and AMM through the MC$\chi$SB mechanism, so the field-dependence of the condensates is not expandable in powers of B, and hence can never be linear.  

\section{Concluding Remarks}

In this paper we reconsidered the effect of an applied magnetic field on the chiral phase transition of a QCD-inspired theory described by a one-flavor NJL model with interactions channels consistent with the QCD symmetries in a magnetic field. With this purpose we worked out the Fierz identities that can be derived from one-gluon exchange interactions in a system where part of the rotational symmetry has been broken explicitly by the external magnetic field.

Using the NJL model with extra tensor channels, we showed that the phase with broken chiral symmetry is characterized by a spin condensate and the conventional chiral condensate.  In the presence of a magnetic field no solution exists with $\langle \overline{\psi}\psi\rangle\neq 0$ and $\langle \overline{\psi}\Sigma^3\psi\rangle= 0$, and vice versa. To understand the genesis of the new condensate, we should take into account that, since the pairs are formed by a particle and an antiparticle with opposite spins and charges, they have their magnetic moments pointing in the same direction. Under an applied magnetic field, the magnetic moments of the pairs orient in the field direction giving rise to an overall MM of the ground state that is equivalent to a nonzero expectation value of $\langle \overline{\psi}\Sigma^3\psi\rangle$. The new condensate dresses the quasiparticles with a dynamical AMM, as reflected in the way the AMM parameter $\xi$ enters in the spectra. The dynamical AMM produces a Zeeman effect in all the quasiparticles with nonzero Landau levels. For LLL quasiparticles, there is no Zeeman splitting because only one spin contributes. However, the effect of $\xi$ in this case is to significantly increase the effective dynamical mass of the LLL quarks, and consequently the critical temperature of the chiral phase transition.

As the quasiparticles will be heavier at large fields, compared to their mass when the spin condensate can be ignored, and since they are charged, the electrical conductivity in this case should be much smaller at strong fields. This will affect the transport properties of this magnetized medium, a topic worthy of more investigation for its potential implications for astrophysics.

Although it seems natural to expect similar results for more general NJL models, it will be important to study the structure of the ground state in the context of more realistic theories including two and three flavors. Another pending task is to explore the density effects, which can be highly nontrivial, judging by what is known to occur in a QCD-like model with inhomogeneous condensates \cite{Quarkyonic}.

It has been recently argued \cite{cuc} that at zero magnetic field the quarks acquire an AMM due to the regular dynamical chiral symmetry breaking mechanism of QCD at supercritical coupling. While this is physically understandable, since once the quarks have a mass, they must also possess an AMM, the magnetic moment of the pairs should not have any net orientation in the absence of a magnetic field. Hence the ground state will have no magnetic order, and the only condensate in this case will be the usual chiral condensate. 

Nevertheless, in the supercritical system, as soon as a magnetic field is present, the alignment of the pairs' magnetic moments will occur and there will be a nonzero expectation value of the system's MM. If the results of the present work give us any indication of the behavior expected in this supercritical regime within the NJL model in a magnetic field, one would expect that this MM will contribute to significantly increasing the critical temperature, as it did in the subcritical case. However, this is not what is found within lattice QCD and one needs to understand why. We expect to explore this important question in the near future.


\begin{acknowledgments}
The work of E. J. Ferrer and V. de la Incera has been supported in part by DOE Nuclear Theory Grant No. DE-SC0002179 and the work of
M. Quiroz by UTEP-COURI-2012 grant. Ferrer and de la Incera thank the Yukawa Institute of Theoretical Physics, Kyoto, for useful discussions during the YITP workshop YITP-T-13-05 on New Frontiers in QCD, where this work was presented.

\end{acknowledgments}

\appendix
\section{Fierz Identities with Rotational Symmetry Breaking}

As is known, the Fierz identities that are connected to a reordering of the field operators in a given interaction depend on the symmetry of the system \cite{AMM-CS, Buballa}. Specifically, in the case of the one-gluon exchange interaction, which at zero momentum is reduced to
a contact four-particle interaction, the Fierz identities give the prescription to make the transformation
\begin{equation}
({\bar\psi_1 \Gamma^A\psi_2})({\bar\psi_3\Gamma_B\psi_4})\rightarrow ({\bar\psi_1\Gamma^C\psi_4})({\bar\psi_3\Gamma_D\psi_2})
\label{bilinear}
\end{equation}
Here, the spinor indices are suppressed. To find the relation between the matrices $\Gamma^A$, $\Gamma^B$ and $\Gamma^C$, $\Gamma^D$ is precisely the goal of the Fierz identities.

In vacuum, the Dirac structures entering in the bilinears in (\ref{bilinear}) are given by the elements of the Dirac ring
\begin{equation}
\{\Gamma^A\}=\{1, \gamma_5, \gamma^\mu, \gamma_5\gamma^\mu, \sigma^{\mu\nu}\},
\label{basis}
\end{equation}

The presence of a constant and uniform magnetic field breaks the rotational $O(3)$ symmetry. This symmetry breaking implies that the tensor structures of the Dirac ring split in parallel and transverse component to the field direction,
\begin{equation}
\{\widehat{\Gamma}^A\}=\{1, \gamma_5, \gamma^\|, \gamma^\bot, \gamma_5\gamma^\|, \gamma_5\gamma^\bot, \sigma^{\|\|}, \sigma^{\|\bot}, \sigma^{\bot\bot}\},
\label{anisotrotic-basis}
\end{equation}
where for a magnetic field along the $x_3$ direction we are using the index notation $\|=0,3$ and $\bot=1,2$.

Then, the Fierz identities connecting the different elements in (\ref{anisotrotic-basis}) are given by \cite{AMM-CS, Buballa, CC}
\begin{equation}
(\Gamma_A)_{ij}(\Gamma^B)_{kl}=\frac{1}{4^2}\text{Tr}[\Gamma^A\Gamma_C\Gamma_B\Gamma_D](\Gamma^D)_{il}(\Gamma^C)_{kj},
\label{identity}
\end{equation}
with all the lower case spinor indices $i,j,k,l$ running over $0, 1, 2, 3$. Then, the expansion coefficients connecting the Dirac elements are straightforwardly obtained as gamma matrix traces from $\frac{1}{4^2}\text{Tr}[\Gamma^A\Gamma_C\Gamma_B\Gamma_D]$.

For the particle-antiparticle channel the anisotropic Fierz identities in the presence of a constant and uniform magnetic field are
\begin{widetext}
\begin{equation}
\left(
\begin{array}{c}
(1)_{ij}(1)_{kl}\\
(\gamma^\|)_{ij}(\gamma_\|)_{kl}\\
(\gamma^\bot)^{ij}(\gamma_\bot)_{kl}\\
(\sigma^{30})_{ij}(\sigma_{30})_{ij}\\
(\sigma^{\bot\|})_{ij}(\sigma_{\bot\|})_{kl}\\
\frac{1}{2}(\sigma^{\bot\bot})_{ij}(\sigma_{\bot\bot})_{kl}\\
(\gamma^\|\gamma_5)_{ij}(\gamma_\|\gamma_5)_{kl}\\
(\gamma^\bot\gamma_5)_{ij}(\gamma_\bot\gamma_5)_{kl}\\
(i\gamma_5)_{ij}(i\gamma_5)_{kl}
\end{array}\right)=\left(
\begin{array}{rrrrrrrrrrrr}
\frac{1}{4} & \frac{1}{4}  & \frac{1}{4} & \frac{1}{4} & \frac{1}{4} & \frac{1}{4} & -\frac{1}{4} & -\frac{1}{4} & -\frac{1}{4} \\
\frac{1}{2} & 0  & -\frac{1}{2}& -\frac{1}{2} & 0 & \frac{1}{2} & 0& -\frac{1}{2}  & \frac{1}{2}\\
\frac{1}{2} & -\frac{1}{2}  & 0 & \frac{1}{2} & 0 & -\frac{1}{2} & -\frac{1}{2} & 0 & \frac{1}{2}\\
\frac{1}{4} & -\frac{1}{4}  & \frac{1}{4} & \frac{1}{4} & -\frac{1}{4} & \frac{1}{4} & \frac{1}{4} & -\frac{1}{4} & -\frac{1}{4}\\
1 & 0  & 0 & -1 & 0 & -1 & 0 & 0 & -1\\
\frac{1}{4} & \frac{1}{4}  & -\frac{1}{4} & \frac{1}{4} & -\frac{1}{2} & \frac{1}{4} & -\frac{1}{4} & \frac{1}{4} & -\frac{1}{4}\\
-\frac{1}{2} & 0  & -\frac{1}{2} & \frac{1}{2} & 0 & -\frac{1}{2} & 0 & -\frac{1}{2} & -\frac{1}{2} \\
-\frac{1}{2} & -\frac{1}{2}  & 0 & -\frac{1}{2} & 0 & \frac{1}{2} & -\frac{1}{2} & 0 & -\frac{1}{2}\\
-\frac{1}{4} & \frac{1}{4}  & \frac{1}{4} & -\frac{1}{4} & -\frac{1}{4} & -\frac{1}{4} & -\frac{1}{4} & -\frac{1}{4} & \frac{1}{4}\\
\end{array}\right)\left(
\begin{array}{c}
(1)_{ij}(1)_{kl}\\
(\gamma^\|)_{ij}(\gamma_\|)_{kl}\\
(\gamma^\bot)^{ij}(\gamma_\bot)_{kl}\\
(\sigma^{30})_{ij}(\sigma_{30})_{ij}\\
(\sigma^{\bot\|})_{ij}(\sigma_{\bot\|})_{kl}\\
\frac{1}{2}(\sigma^{\bot\bot})_{ij}(\sigma_{\bot\bot})_{kl}\\
(\gamma^\|\gamma_5)_{ij}(\gamma_\|\gamma_5)_{kl}\\
(\gamma^\bot\gamma_5)_{ij}(\gamma_\bot\gamma_5)_{kl}\\
(i\gamma_5)_{ij}(i\gamma_5)_{kl}
\end{array}\right),
\end{equation}
\end{widetext}
where repeated indexes denote summation.

\end{document}